\documentclass[prl,twocolumn,showpacs,floatfix,amsbsy,amsbsy,superscriptaddress]{revtex4-1}
\usepackage{epsfig,color}
\usepackage{amsmath}
\usepackage{amsfonts}
\usepackage{color}
\usepackage{latexsym}
\usepackage{hyperref}
\usepackage{blindtext}
\begin{document}
\title{Spin and Charge Signatures of Topological Superconductivity in Rashba Nanowires}

\author{Pawe\l{} Szumniak}
\affiliation{Department of Physics, University of Basel, Klingelbergstrasse 82, 4056 Basel, Switzerland}
\affiliation{AGH University of Science and Technology, Faculty of
Physics and Applied Computer Science,\\
al. Mickiewicza 30, 30-059 Krak\'ow, Poland}
\author{Denis Chevallier}
\affiliation{Department of Physics, University of Basel, Klingelbergstrasse 82, 4056 Basel, Switzerland}
\author{Daniel Loss}
\affiliation{Department of Physics, University of Basel, Klingelbergstrasse 82, 4056 Basel, Switzerland}
\author{Jelena Klinovaja}
\affiliation{Department of Physics, University of Basel, Klingelbergstrasse 82, 4056 Basel, Switzerland}
\date{\today}
\begin{abstract}
We consider a Rashba nanowire with proximity gap which can be brought into the topological phase by tuning external magnetic field or chemical potential. We study spin and charge of the  bulk quasiparticle states when passing through the topological transition for open and closed systems.We show, analytically and numerically, that the  spin of bulk states around the topological gap reverses its sign when crossing the transition due to band inversion, independent of the presence of  Majorana fermions in the system. 
This  spin reversal can be considered as a bulk signature of  topological superconductivity  that can be accessed experimentally.
We find a similar behaviour for the charge of the bulk quasiparticle states, also exhibiting a sign reversal at the transition. We show that these signatures are robust against random static disorder.
\end{abstract}

\maketitle

\emph{Introduction.} 
Topological phases of condensed matter systems~\cite{Hassan_Kane,Bernevig_book} have attracted a lot of attention over many years due to their high promise for applications such as topological quantum computation~\cite{Kitaev,Stern_RMP}. One of the hallmarks of such phases, in particular of topological superconductivity, are zero-energy modes such as Majorana fermions (MF) that emerge at the edges of the system. 
Various candidate materials can host such topological states \cite{Igor,Manisha2,Setiawan, Fu,MF_Sato,MF_Sarma,MF_Oreg,alicea_majoranas_2010,potter_majoranas_2011,
Klinovaja_CNT,Pascal,Fra,Rotating_field,Ali,RKKY_Basel,RKKY_Simon,Carlos} but one of the most promising platforms are semiconducting nanowires of InAs or InSb material, with strong Rashba spin orbit interaction (SOI), subjected to an external magnetic field and in proximity to an $s$-wave superconductor  \cite{Alicea,Beenakker}. Experimental evidence has been reported for topological phases in such 
wires~\cite{Mourik,Marcus,Heiblum,Rokhinson,Xu,Kuemmeth_Nature2016,Frolov,Deng_Marcus_Science2017} as well as  in magnetic atomic chains on superconducting substrates \cite{Franke,Yazdani,Meyer}. However, most of the work so far has focused on the detection 
of the MFs in these nanowires and not on their bulk properties. This is quite surprising given the fact that the unambiguous identification of MFs 
from transport data alone can be challenging~\cite{DeFranceschi2,DeFranceschi,Liu2012,Atland2012,Pikulin2012,Diego,Stanescu_2017,Chris,Manisha}.
It is thus of great interest to look for alternative signatures of topological phases and to address the question  how the bulk states change when passing from trivial to topological phase and if these changes appear in physically observable quantities.

In this work, we show that the phase of a topological superconductor can be monitored by  bulk states, in particular by certain spin and charge degrees of freedom.  Quite remarkably, we find that the sign of the spin component along the magnetic field reverses for low-momentum states close to the Fermi level when the system passes through the phase transition, and similarly for the charge of such bulk states. This sign reversal is a direct consequence of the band inversion at the transition point and is directly accessible by spin- and energy resolved measurements. Another remarkable feature is that this signature is independent of boundary effects and thus unrelated to the presence of MFs. To demonstrate these findings we perform analytical and numerical calculations for both closed and open systems and for various parameter regimes which are relevant for InAs or InSb nanowires used in recent experiments~\cite{Frolov,Mourik,Marcus,Heiblum,Rokhinson,Xu,Kuemmeth_Nature2016,Deng_Marcus_Science2017}. We also demonstrate that these effects are robust against static random disorder.
 
\begin{figure}[t]
\includegraphics[width=8.6cm]{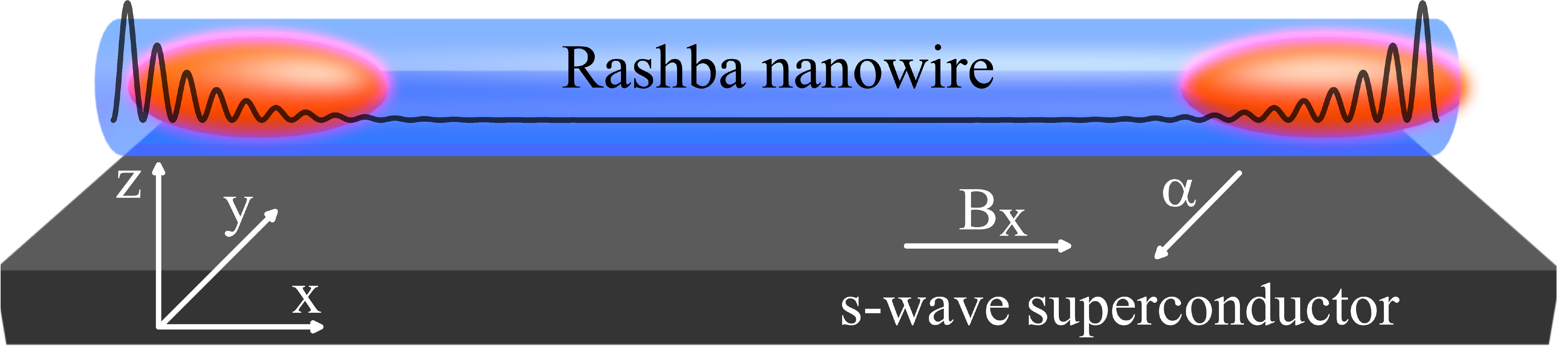}
\caption{A semiconducting nanowire proximity-coupled to an $s$-wave superconductor.  The Rashba SOI vector ${\bf \alpha}$ points in the $y$-direction and an external magnetic field ${ B_x}$ is applied in the $x$-direction. 
In the topological phase, spin- and chargeless MFs (orange ovals) are localized at the ends, with corresponding probability density 
indicated by black lines.
}
 \label{fig:Setup}
\end{figure}

{\it Model.} We consider a one-dimensional Rashba nanowire aligned along the $x$-axis and placed on top of an $s$-wave superconductor in the presence of an external magnetic field applied along the nanowire axis (see Fig.~\ref{fig:Setup}). The system can be modeled by the tight-binding Hamiltonian
\begin{align}
   &{H}=\sum_{j=1}^{N-1}[{\Psi}_{j+1}^\dag(-t-i\tilde{\alpha}\sigma_y)\tau_z{\Psi}_j+{\text {H.c.}}]\nonumber\\
	  &\hspace{40pt}+\sum_{j=1}^N{\Psi}_j^\dag[(2t-{\mu})\tau_z+\Delta_{sc}\tau_x+\Delta_{Z}\sigma_x]{\Psi}_j \label{hamiltonian_position},
\end{align}
where ${\Psi}_j=({c}_{j \uparrow}, {c}_{j \downarrow}, {{c}^{\dag}_{j \downarrow}}, {-{c}^{\dag}_{j \uparrow}})^T$ is given in standard Nambu representation.  The creation operator ${c}^{\dag}_{j \sigma} $ acts on an electron with spin $\sigma$ located at site $j$ in a chain of $N$ sites with lattice constant $a$.
 The Zeeman splitting 
 $2\Delta_Z=g\mu_B B_x$ is determined by the $g$-factor and by the strength of the external magnetic field $B_x$.  The superconducting pairing term  $\Delta_{sc}$ is induced in the nanowire via proximity effect by the $s$-wave superconductor.  The chemical potential of the nanowire $\mu$ is calculated from the SOI energy  and  $t$ is the hopping amplitude. The Pauli matrices $\sigma_i$ ($\tau_i$) act on spin (particle-hole) space and $\tilde{\alpha}$ is the spin-flip hopping amplitude used to model the Rashba SOI. By diagonalizing numerically ${H}$, we find the spectrum $E_n$ and corresponding wavefunctions $\Phi_n (j)$ labeled by the index $n=1,...,4N$.
 
\begin{figure}[t]
\centering
\includegraphics[width=8.6cm]{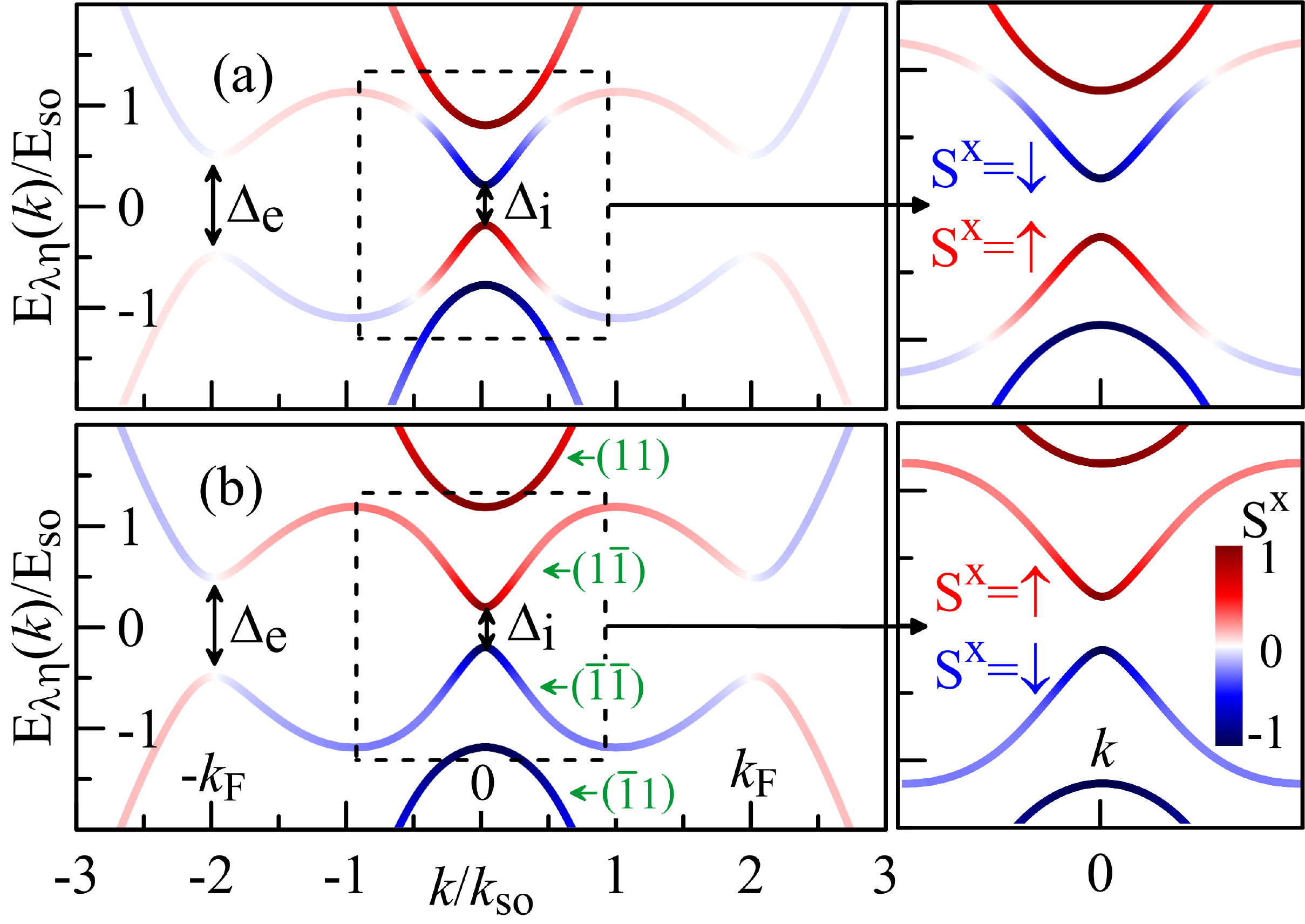}
\caption{The spectrum of Eq.  ($\ref{hamiltonian_momentum_open}$)  in the trivial (a) and topological (b) phases for closed systems. The blue/red color of dispersion lines $E_{\lambda\eta} (k)$ indicates the negative/positive sign of the spin component  $S^x_{\lambda\eta}(k)$  
for states of a given $\lambda\eta$-band. 
Around $k=0$, the spin   $S^x_{\lambda\bar1}(k)$ of  states with energies closest to the Fermi level $\mu=0$ in the trivial phase is opposite to the one in the topological phase. The insets show the spectrum around $k=0$ where such a  sign reversal of $S^x_{\lambda\bar1}(k)$ 
occurs. We used the following parameters: $\mu=0$, $\Delta_{sc}=0.5E_{so}$ and $\Delta_{Z}=0.3E_{so} $ ($\Delta_{Z}=0.7E_{so}$) in the trivial (topological) phase, so that the interior gap $\Delta_i=2|\Delta_{Z}-\Delta_{sc}|$ remains the same.}
\label{fig:band_structure_Sx}
\end{figure}

In order to study analytically the bulk states of the system,  we also write ${H}$ in  momentum space.  By imposing periodic boundary conditions, we can introduce the momentum $k$,  ${c}_{j \sigma}=\sum_j {c}_{k \sigma} e^{-ijka}/\sqrt{N}$, and ${\Psi}_k=({c}_{k \uparrow}, {c}_{k \downarrow}, {c}_{-k \downarrow}^{\dag}, -{c}_{-k \uparrow}^{\dag})^T$. We then obtain  ${H}=\sum_k{\Psi}_k^\dag{\mathcal H}(k){\Psi}_k$ with
\begin{align}
 &  {\mathcal H}(k)=\left[2t-2t\cos(ka)-\mu+2\tilde{\alpha}\sin(ka)\sigma_y\right]	\tau_z \nonumber\\
    &\hspace{100pt}+\Delta_{sc}\tau_x+\Delta_Z\sigma_x. \label{hamiltonian_momentum_tb}
\end{align} 
In the continuum limit $ka\ll 1$ \cite{composite_majorana},
we get 
\begin{equation}\label{hamiltonian_momentum_open}
    {\mathcal H}(k)=\left(\frac{\hbar^2k^2}{2m} -\mu+\alpha k \sigma_y\right)\tau_z+\Delta_{sc}\tau_x+\Delta_{Z}\sigma_x.
\end{equation}
The correspondence between the tight-binding and continuum model is then given by $t=\hbar^2/(2ma^2)$, where $m$ is an effective electron mass \cite{Diego}. The spin-flip hopping amplitude  $\tilde{\alpha}$ is related to the SOI strength by $\tilde{\alpha}=\alpha/2a$. The corresponding SOI energy (momentum) is defined as $E_{so}=\tilde{\alpha}^2/t\equiv m \alpha^2/2 \hbar^2$  ($k_{so}=m\alpha/\hbar^2$).
By diagonalizing $ {\mathcal H}(k)$ [see Eq.  ($\ref{hamiltonian_momentum_tb}$) or ($\ref{hamiltonian_momentum_open}$)], we arrive at analytical expressions for the eigenvalues  $E_{\lambda\eta}(k)$ and corresponding eigenstates $\Phi_{\lambda\eta}(k)$ (see SM~\cite{SM}).
In total, there are four bands, labeled by  $\lambda$ and $\eta$, where $\lambda=1$ ($\lambda=\bar 1$) labels bands with positive (negative) energy and $\eta=\bar 1$ the bands closest to the Fermi level,  see Fig. \ref{fig:band_structure_Sx}. 

The lowest band $\eta=\bar 1$ has gaps at $k=0$, which we call the interior gap $\Delta_i=2|\Delta_z-\sqrt{\mu^2+\Delta_{sc}^2}|$, and at finite Fermi points $k=\pm k_F$, which we call the exterior gap $\Delta_e=2|E_{\lambda\bar 1}(k_F)|$.
The central quantities of interest are the
spin ${\bf S}_{n}$ [${\bf S}_{\lambda\eta}(k)$] and the charge $Q_n$ [$Q_{\lambda\eta}(k)$] of  the bulk quasiparticles states at given energy $E_n$ [$E_{\lambda\eta}(k)$],
defined in the tight-binding (continuum) model as 
\begin{align}
&{\bf S}_{n}~=~\sum_{j=1}^N\Phi_n^\dag(j){\boldsymbol\sigma}\Phi_n(j), \label{spin_formula}\\
&{\bf S}_{\lambda\eta}(k)~=~\Phi_{\lambda\eta}^\dag(k){\boldsymbol\sigma}\Phi_{\lambda\eta}(k),\label{spin_formula_k}\\
&Q_n~=~-\sum_{j=1}^N\Phi_n^\dag(j)\tau_z\Phi_n(j),\label{charge_formula} \\
&Q_{\lambda\eta}(k)~=~-\Phi_{\lambda\eta}^\dag(k)\tau_z\Phi_{\lambda\eta}(k).\label{charge_formula_k}
\end{align}
Here, the spin and charge are  measured in units of $\hbar/2$ and electron charge $|e|$, respectively.  We note that in contrast to previous works \cite{Annica,Bena}, our definitions of spin and charge involve all four components of the wavefunction.
Due to the periodic boundary conditions the system describes a closed ring and no MF can occur (for open systems, see below).

{\it Spin and charge inversion at the topological phase transition.} 
We focus now on the spin and charge of bulk states of the nanowire in the trivial ($\Delta_Z^2<\mu^2+\Delta^2_{sc}$) and in the topological ($\Delta_Z^2>\mu^2+\Delta^2_{sc}$) phases, see Fig.~\ref{fig:band_structure_Sx}.
The most interesting behavior occurs close to $k=0$, where the topological  phase transition takes place as $\Delta_i=0$ for the $\eta=\bar 1$ band. Quite remarkably, we observe a sign reversal of the spin component along the magnetic field, $S^x_{\lambda \bar1 }(k)$, when the system is tuned from trivial to topological phase.
In Fig. \ref{fig:band_structure_Sx}(a), the system is shown in the trivial phase where $S^x_{1\bar 1}(k)$ [$S^x_{\bar1\bar 1}(k)$] around $k=0$ is negative (positive) for the electron (hole)  $\eta=\bar 1$ band, while the sign reverses when the system is tuned into the topological phase by changing the magnetic field, see Fig. \ref{fig:band_structure_Sx}(b).
This change of sign is a direct consequence of the band inversion associated with the topological phase transition. Consequently, by measuring the spin component $S^x$ along the field $B_x$, one can determine whether the system is in the topological or trivial phase, even in the absence of any MFs.
 This  finding opens up new experimental perspectives to identify topological superconductivity by measuring bulk state properties close to the Fermi level.  

We note that there is also a residual spin component along the SOI axis $S^y(k)$, the sign of which, however, is the same  both in the topological and trivial phase and thus cannot be used to distinguish phases. Moreover, due to the symmetry of the system, the spin projection $S^z(k)$ is always zero. In the Supplemental Material (SM) \cite{SM}, we provide the analysis of all spin components $S^i(k)$ 
as a function of momentum $k$.
We finally note that 
similar behavior as for spin is found also for the quasiparticle charge as shown in the SM. Indeed,
close to $k=0$ and for the negative value of $\mu$ in the topological (trivial) phase, $Q_{1 \bar 1}(k)$ is positive (negative) while $Q_{\bar 1 \bar 1}(k)$ is  negative (positive). For positive values of $\mu$ the situation is opposite.
Again, this sign reversal can be used as a detection tool for topological superconductivity, independent of MFs. In the SM \cite{SM}, we also demonstrate that our results hold for nanowires with several subbands.

\begin{figure}[t]
\centering
\includegraphics[width=8.6cm]{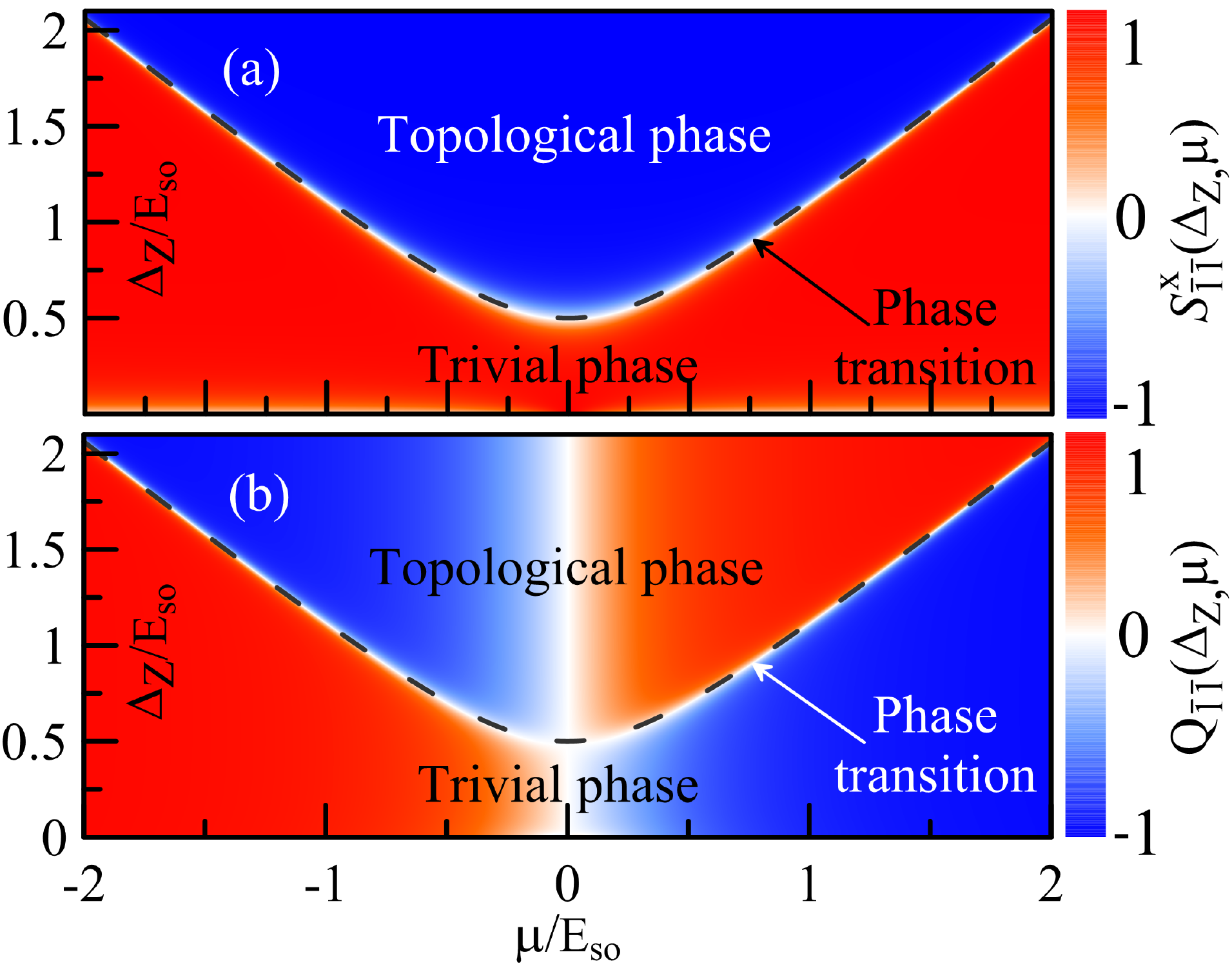}
\caption{ 
Phase diagram as function of chemical potential $\mu$ and magnetic field $\Delta_Z$ by measuring (a) the $x$ component of the quasiparticle spin $S^x_{\bar1\bar1}$ and (b) the quasiparticle charge $Q_{\bar1\bar1}$ at fixed momentum $k=0.05k_{so}$ and fixed superconducting pairing $\Delta_{sc}=0.5E_{so}$. 
 Indeed, the phase boundary at the topological  transition line $\Delta_Z=\sqrt{\Delta_{sc}^2+\mu^2}$ (dashed line) is very well visible.
A similar phase diagram is found for the states above $\mu$ ($S^x_{1\bar1}$ and $Q_{1\bar1}$), differing only in a global minus sign due to  particle-hole symmetry.}
  \label{fig:phase_diagram}
\end{figure}

\emph{Phase Diagram.}
To test if the spin $S^x_{\lambda\bar1}$ and charge $Q_{\lambda\bar1}$ of the bulk states with momentum close to $k=0$ allows one to distinguish reliably between trivial and topological phase, we determine the phase diagram as a function of magnetic field $\Delta_Z$ and chemical potential $\mu$ at fixed momentum, see Fig. \ref{fig:phase_diagram}, again for the closed system without MFs.  The results are obtained for the bulk states from the $\bar 1 \bar 1$-band, using 
 Eq.  ($\ref{hamiltonian_momentum_open}$).
The phase transition at $\Delta_Z^2=\Delta_{sc}^2+\mu^2$ is clearly indicated by the reversal of signs of both the spin $S^x_{\lambda\bar1}$ and charge $Q_{\lambda\bar1}$.  Moreover, the boundary separating the two phases is sharp.
We note that the charge reverses its sign at  $\mu=0$. In contrast to the topological phase transition, this phase boundary is smooth.

The analytical expressions for charge and spin of the bulk states, obtained from Eqs. (\ref{spin_formula_k}) and (\ref{charge_formula_k}), are too lengthy to be shown here.  However, since we are mainly interested in the features around $k=0$, we can expand these formulas for small momenta away from the phase transition (for simplicity, we also put $\mu=0$).
 For the $\lambda\bar1$-bands we get in leading order,
\begin{align}
&S^x_{\lambda\bar 1}(k)=\lambda\textrm{sign}(\Delta_Z-\Delta_{sc})\Big[1-\frac{(\alpha k)^2}{2(\Delta_Z-\Delta_{sc})^2}\Big] \label{spin_analytic}\, ,\\
&Q_{\lambda\bar 1}(k)=\lambda\textrm{sign}(\Delta_Z-\Delta_{sc})\frac{\hbar^2k^2}{2m\Delta_{sc}}\label{charge_analytic}\, .\\\notag
\end{align}
We can clearly see that around $k=0$ the sign of $S^x_{\lambda\bar 1}$
is proportional to the sign of the topological gap $\Delta_Z-\Delta_{sc}$.
Thus, one can consider $S^x_{\lambda\bar 1}(k)$  as an order parameter that distinguishes between topological and trivial phases.  One also notices that the sign of the quasiparticle charge is proportional to the sign of the topological gap, however it changes only quadratically in $k$.

\begin{figure}[b]
\includegraphics[width=8.6cm]{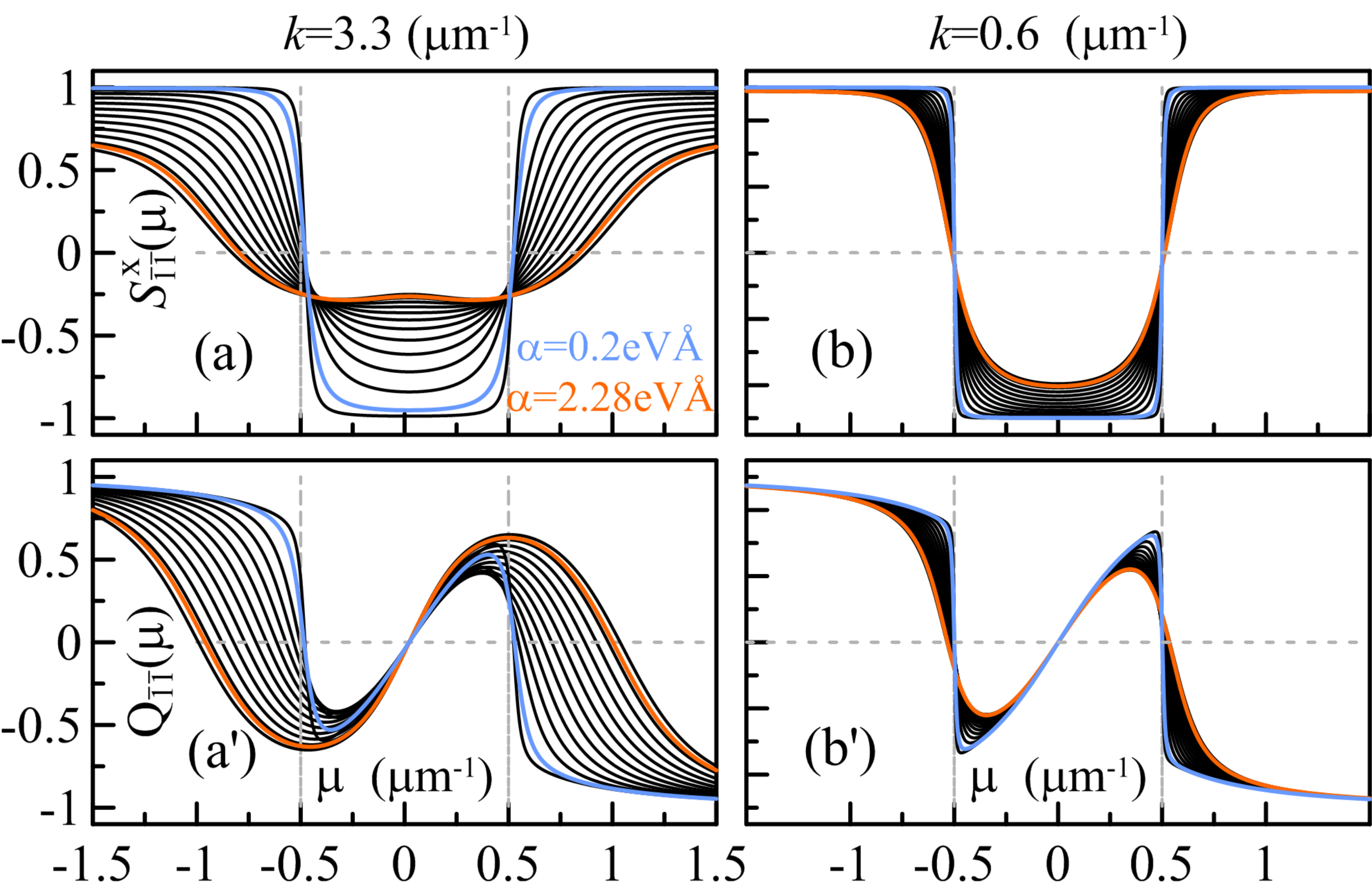}
\caption{
Spin projection $S^x_{\bar1\bar 1}$ [panels (a,b)] and charge $Q_{\bar1\bar 1}$  [panels (a',b')] of bulk states as a function of the chemical potential $\mu$ for two values of momenta (a,a') $k=3.3$~$\mu$m$^{-1}$ and (b,b') $k=0.6$~$\mu$m$^{-1}$ for various values of the SOI strength: $\alpha=0.1, 0.2, 0.4 ...,2.2, 2.28, 2.4$eV\AA. 
The parameters of the system are $\Delta_{sc}=0.5$~meV and $\Delta_Z=0.7$~meV.
The sign of charge and spin reverses as the system undergoes the topologiocal phase transition at $\mu=\pm 0.5$~meV (denoted by gray dashed vertical lines). The boundary between phases is most pronounced for the states close to $k=0$ and almost independent of the SOI.}
  \label{fig:Q_S(a_mu)_without_map}
\end{figure}

So far we have studied systems with strong SOI, which is generally believed to be the case for InSb or InAs nanowire \cite{Mourik}. However, the sharpness of the boundary between two phases determined by $S^x_{\lambda \bar 1}$ depends on the strength of the SOI as seen from Eq. (\ref{spin_analytic}). To understand this dependence better, we study the evolution of $S^x_{\lambda \bar 1}$ and $Q_{\lambda \bar 1}$ as a function of $\mu$ for two different values of $k$, see Fig. \ref{fig:Q_S(a_mu)_without_map}. Generally, we observe that the boundary between the topological and  trivial phase is sharper for small values of $k$ and almost insensitive to the SOI strength. To conclude, the reversal of the sign of  $Q_{\lambda \bar 1}$ and  $S^x_{\lambda \bar 1}$  is well visible at small $k$ for all values of SOI.

\begin{figure}[t]
\centering
\includegraphics[width=8.6cm]{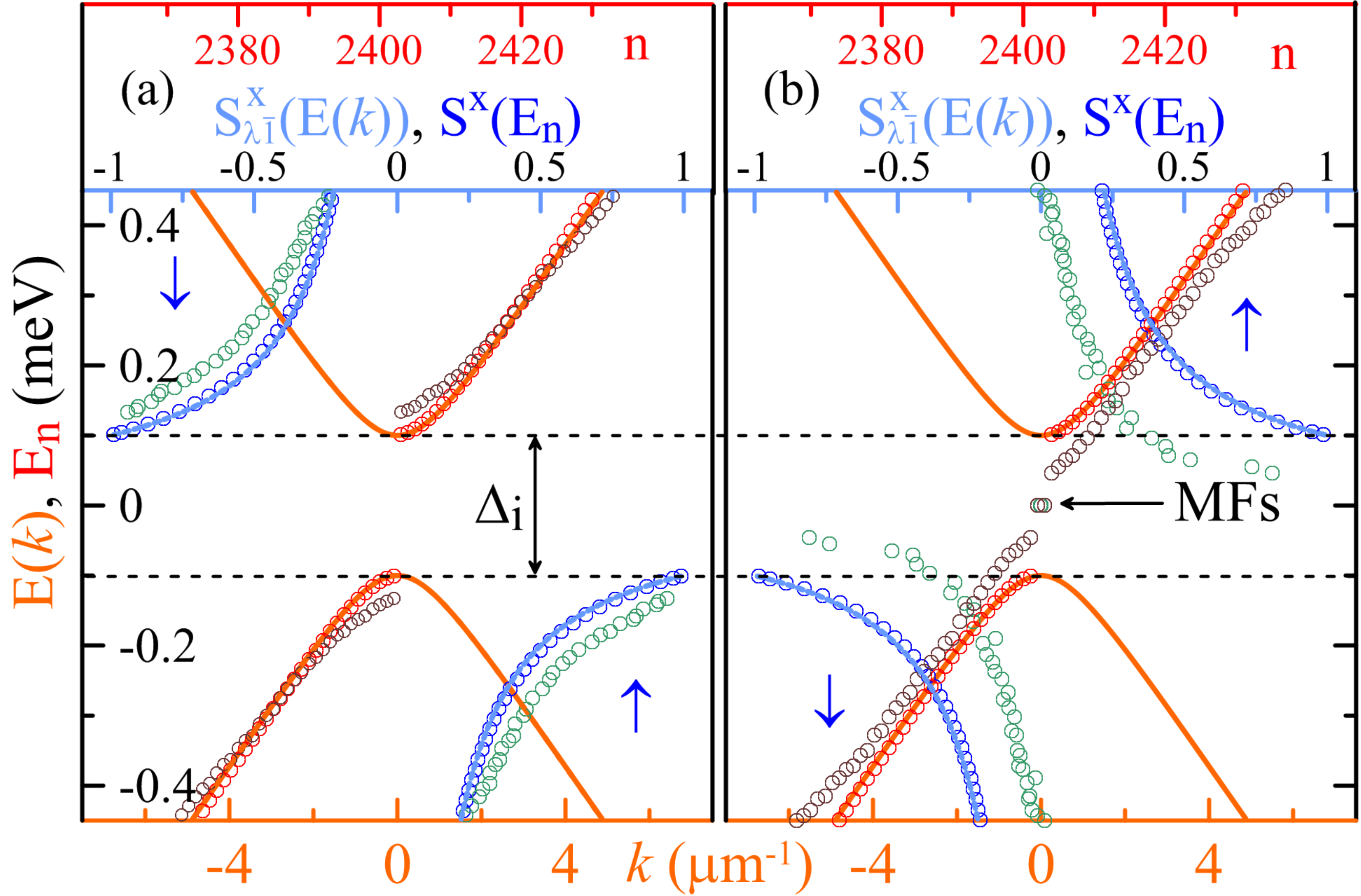}
\caption{The energy spectrum and spin $S^x$ in the trivial (a) and  topological (b) phase for open and closed systems.  We find excellent agreement between $E_n$ (red circles) and $S^x(E_n)=S^x_n$ (blue circles) obtained from the tight-binding model for open systems supporting MFs [see Eq. (\ref{hamiltonian_position})] and $E_{\lambda\bar1}(k)$ (orange solid line) and $S^x_{\lambda\bar1}(k)$ obtained for closed systems  without MFs [see Eq. (\ref{hamiltonian_momentum_tb})].  The topological phase transition is clearly indicated by the sign reversal of $S^x$ also in  open systems.
The trivial phase is plotted for $\Delta_Z=0.4$~meV while the topological one for $\Delta_Z=0.6$~meV. The other parameters are fixed as $\Delta_{sc}=0.5$~meV, $\mu=0$, 
$\alpha=0.9$~eV\AA  ($E_{so}\approx1$~meV),
$N=1200$ and $m=0.015m_e$, with $m_e$ being the bare electron mass, for InSb nanowires ($t=10$~meV, $a=15$~nm). 
In addition, we consider a random on-site disorder potential of  strength $|\delta \mu_j| < 1$meV.
The spin $S^x(E_n)$ (green circles) corresponding to $E_n$ (brown circles) of the disordered nanowire undergoes the same reversal of  sign as  in the clean case, demonstrating its robustness.  }
  \label{fig:spectrum_numerics}
\end{figure}

{\it Open systems with MFs.}  So far we have considered closed systems not supporting MFs.
In  realistic setups, nanowires are open with finite length and momentum is not a good quantum number. 
In addition,  finite systems in the topological phase host MFs at the wire ends and it is apriori not clear if their presence does not mask the signatures of topological phase transition found for closed systems above.
 Thus, we focus now on such finite systems
using realistic parameters \cite{Mourik,Kammhuber2017}. First, we compare results obtained from Eq. (\ref{hamiltonian_momentum_tb}) for closed systems with periodic boundary conditions with the ones obtained from tight-binding calculations using Eq. (\ref{hamiltonian_position})  for open systems with vanishing boundary conditions, see  Fig. \ref{fig:spectrum_numerics}. Our numerical simulations give excellent agreement between the two models for parameters for which $\Delta_{i}<\Delta_{e}$. In the opposite regime, the exterior and interior branches are mixed, thus, one needs to involve momentum-resolved measurements. 
% for energies up to $\Delta_e/2$. 
 Thus,
 the spin component $S^x$ and its reversal can serve as a detection tool also in open systems.  We obtain  similar results also for the quasiparticle charge $Q_n$, see SM~\cite{SM}. All this confirms that the topological phase transition of the bulk  can be detected in the same way. The sign reversal of spin and charge  does not depend on boundaries of the system, and is thus  independent of the presence of MFs. This provides an advantage over detecting the topological phase via the presence of MFs, which could either leak into the lead \cite{composite_majorana} or be masked by disorder effects \cite{Pikulin2012,Diego,Liu2012,Atland2012}. To show that our results are robust against disorder, we add random on-site fluctuations to $\mu$, i.e. set $\mu_i=\mu + \delta \mu_i$ in Eq. (1). We find that even for disorder strengths exceeding the proximity gap $\Delta_{sc}$, the reversal of sign in spin is hardly affected, see Fig.~\ref{fig:spectrum_numerics}. In the SM \cite{SM}, we study effects of static potential and magnetic disorder on the charge and spin signature in more detail.  Again, we conclude that the proposed signature could be used to characterize the topological phase transition.

\emph{Conclusions.}  
We have shown that the topological phase transition in Rashba nanowires is characterized by a sign reversal of the spin component along the magnetic field and of the charge of bulk states with momenta close to $k=0$. Importantly, these findings are independent of the presence of MFs and rely only on bulk properties of the system. 
The boundary between phases is quite sharp but depends on the parameters of the system such as the SOI strength. These results open a way for mapping the phase diagram of the Rashba nanowire and bring a clear signature of the topological phase transition. Two types of experiments could be carried out to detect the spin or charge reversal at the topological phase transition point. The first one is based on  spin-polarized STM spectroscopy which allows one to inject a current in the lowest bands \cite{Yazdani,Franke,Meyer,STM}. Depending on the polarization of the STM probe, a current will flow or not in the trivial phase and the opposite situation will occur in the topological phase.  The second possibility is to couple the nanowire to a quantum dot~\cite{Carlos,dot1,DeFranceschi2,Kouwenhoven_2016,Deng_Marcus_Science2017} which then can be used for 
energy-selective spin read-out~\cite{HansonRMP_2007}.

{\it Acknowledgements.} We acknowledge helpful discussions with Silas Hoffman, Marcel Serina, and Andreas Baumgartner. This work was support by the Swiss NSF and  NCCR QSIT.

\newpage

\onecolumngrid

\bigskip 

\begin{center}
\large{\bf Supplemental Material for ``Spin and Charge Signatures of Topological Superconductivity in Rashba Nanowires'' \\}
\end{center}
\begin{center}
Pawe\l{} Szumniak$^{1,2}$, Denis Chevallier$^{1}$, Daniel Loss$^{1}$ and Jelena Klinovaja$^{1}$\\
$^{1}$ {\it Department of Physics, University of Basel, Klingelbergstrasse 82, CH-4056 Basel, Switzerland}\\
$^{2}$ {\it AGH University of Science and Technology, Faculty of
Physics and Applied Computer Science,\\
al. Mickiewicza 30, 30-059 Krak\'ow, Poland}
\end{center}

\section{Hamiltonian in momentum space for periodic boundary conditions (closed system)}\label{sup_analytic}

We study the bulk states of the system analytically in the momentum space $k$ by imposing periodic boundary conditions, which corresponds to a closed ring without boundary and, thus, to the system without MFs. The operators in the momentum space $c_{k \sigma}$ are defined to the operators in the real space $    c_{j \sigma}$ via Fourier transformation as
\begin{align}
    c_{j \sigma}=\frac{1}{\sqrt{N}}\sum_{j=1}^N c_{k \sigma} e^{-ijka},
\end{align}
Hence, we can write Hamiltonian $H~=~\sum_k\Psi_k^\dag\mathcal H(k)\Psi_k$  in the momentum space as
\begin{equation*}
\mathcal H(k)=
\left( \begin{array}{cccc}
-2t\cos(ka)-\tilde{\mu} & \Delta_Z-2i\tilde{\alpha}\sin(ka) & \Delta_{sc} & 0\\
\Delta_Z+2i\tilde{\alpha}\sin(ka) & -2t\cos(ka)-\tilde{\mu} & 0 & \Delta_{sc}\\
\Delta_{sc} & 0 & 2t\cos(ka)+\tilde{\mu} & \Delta_Z+2i\tilde{\alpha}\sin(ka)\\
0 & \Delta_{sc} & \Delta_Z-2i\tilde{\alpha}\sin(ka) & 2t\cos(ka)+\tilde{\mu}
\end{array}\right),
\end{equation*}
where we define ${\Psi}_k=({c}_{k \uparrow}, {c}_{k \downarrow}, {c}_{-k \downarrow}^{\dag}, -{c}_{-k \uparrow}^\dag)$ as well as $\tilde \mu= \mu -2t$. The other parameters are defined in the main text. The spectrum consists of four bands with energies given by
\begin{equation}
 E_{\lambda\eta}(k)=\lambda\left[\left[\tilde{\mu}+2t\cos(ka)\right]^2+\left[2\tilde{\alpha}\sin(ka)\right]^2+\Delta_Z^2+\Delta_{sc}^2+2\eta\sqrt{\Delta_Z^2\Delta_{sc}^2+[\tilde{\mu}+2t\cos(ka)]^2\left( [2\tilde{\alpha}\sin(ka)]^2+\Delta_Z^2\right)}\right]^{1/2} \nonumber\\
\end{equation}
where the index $\lambda$ labels electron ($\lambda=1$) and hole ($\lambda=\bar 1$) bands. The corresponding wavefunctions are given by
\begin{equation*}
\Phi_{\lambda\eta}(k)=
\frac{1}{\sqrt{N}}
\left(
\begin{array}{c}
 -i(\eta[E_{\lambda\eta}-\tilde{\mu}-2t\cos(ka)][E_{11}^2-E_{1\bar{1}}^2]/4+[\tilde{\mu}+2t\cos(ka)][(2\tilde{\alpha}\sin(ka))^2+\Delta_Z^2])\\
 -i\Delta_Z\Delta_{sc}^2+(2\tilde{\alpha}\sin(ka)-i\Delta_Z)([\tilde{\mu}+2t\cos(ka)]^2-[\tilde{\mu}+2t\cos(ka)]E_{\lambda\eta}-\eta[E_{11}^2-E_{1\bar{1}}^2]/4) \\
\Delta_{sc}(\Delta_Z[2\tilde{\alpha}\sin(ka)-i\Delta_Z]-i\eta[E_{11}^2-E_{1\bar{1}}^2]/4)\\ 
 \Delta_{sc}(2\tilde{\alpha}\sin(ka)[\tilde{\mu}+2t\cos(ka)]+i\Delta_Z E_{\lambda\eta})\\
\end{array}
\right).
\end{equation*}
where $N$ is the normalization factor such as $|\Phi_{\lambda\eta}(k)|^2=1$.

In the continuum limit $ka\ll 1$ the Hamiltonian is rewritten as $H~=~\sum_k\Psi_k^\dag\mathcal H(k)\Psi_k$, where
\begin{equation*}
\mathcal H(k)=
\left( \begin{array}{cccc}
\hbar^2k^2/2m-\mu & \Delta_Z-i\alpha k & \Delta_{sc} & 0\\
\Delta_Z+i\alpha k & \hbar^2k^2/2m-\mu & 0 & \Delta_{sc}\\
\Delta_{sc} & 0 & -\hbar^2k^2/2m+\mu & \Delta_Z+i\alpha k\\
0 & \Delta_{sc} & \Delta_Z-i\alpha k & -\hbar^2k^2/2m+\mu
\end{array}\right).
\end{equation*}
The spectrum consists of four bands with eigenvalues 
\begin{eqnarray}
    E_{\lambda\eta}(k)=\lambda \left [\left(\frac{\hbar^2 k^2}{2m}-\mu\right)^2+(\alpha k)^2+\Delta_Z^2+\Delta_{sc}^2+2\eta\sqrt{\Delta_Z^2\Delta_{sc}^2 + \left(\frac{\hbar^2 k^2}{2m} -\mu \right )^2[(\alpha k)^2+\Delta_Z^2]}\right ]^{1/2},
\end{eqnarray}
and the corresponding eigenstates are given by
\begin{equation}
\Psi_{\lambda\eta}(k)=
\frac{1}{\sqrt{N}}
\begin{pmatrix}
-i [ \eta (E_{\lambda\eta} -\hbar^2k^2/2m+\mu)(E_{11}^2 -E_{1\bar1}^2)/4  +(\hbar^2k^2/2m+\mu) (\alpha^2 k^2 +\Delta_Z^2) ]\\
-i \Delta_Z \Delta_{sc}^2 +(\alpha k - i \Delta_Z )[(\hbar^2 k^2/2m-\mu)^2 - (\hbar^2 k^2/2m-\mu) E_{\lambda \eta }-\eta (E_{11}^2 - E_{1\bar1}^2)/4] \\
\Delta_{sc} [\Delta_Z (\alpha k - i \Delta_Z )- i \eta (E_{11}^2 -E_{1\bar1}^2/4) ] \\
\Delta_{sc} [\alpha k (\hbar^2k^2/2m-\mu)+ i \Delta_Z E_{\lambda  \eta}]
\end{pmatrix}.
\end{equation}

\section{The dependence of the quasiparticles spin  $\bf S$ on momentum (closed system)}\label{Sx_Sy}

\begin{figure}[ht]
\centering
\includegraphics[width=12.5cm]{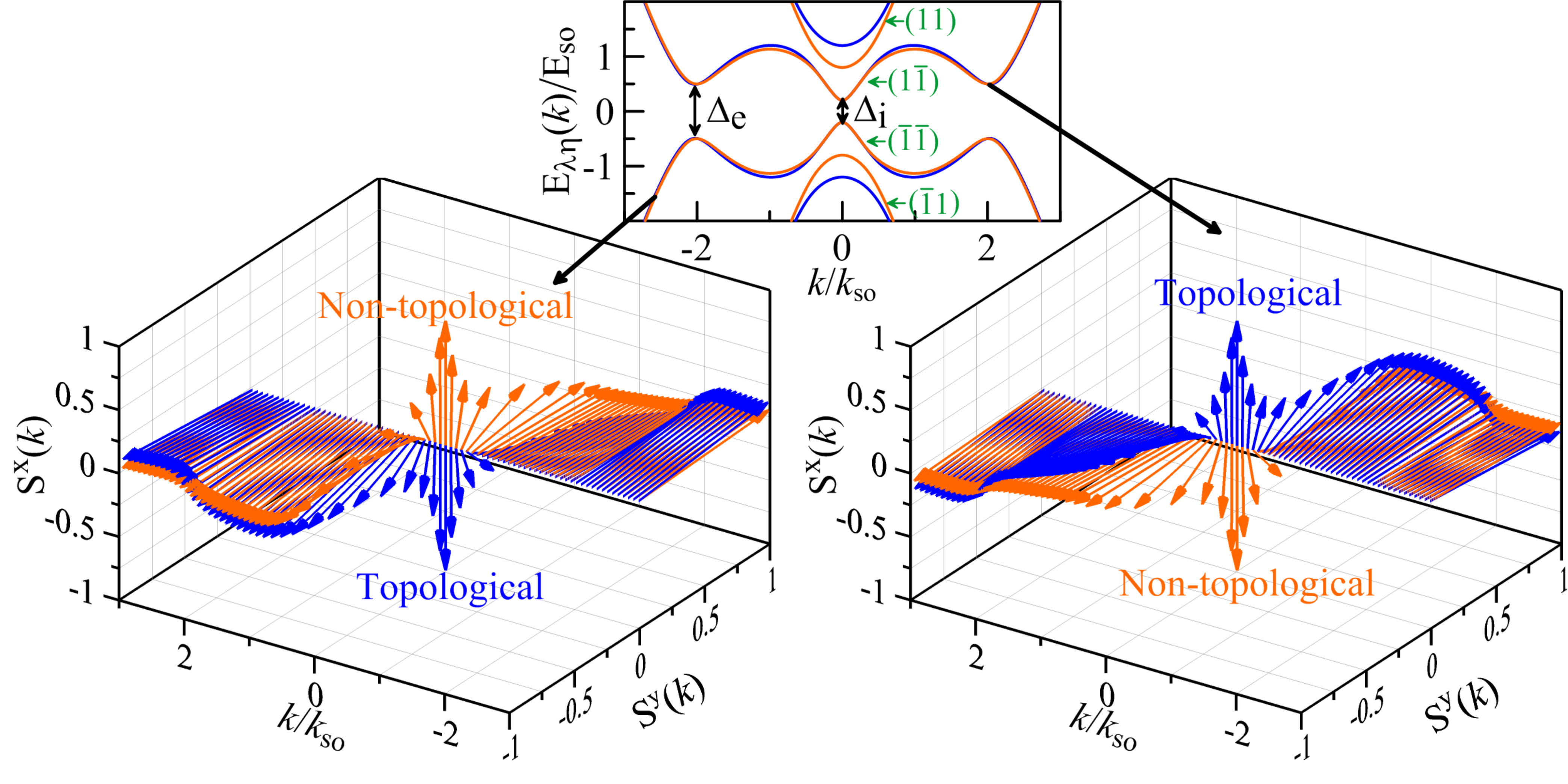}
\caption{The spin projections ${ S}^x_{\lambda \bar1}$ and ${ S}^y_{\lambda \bar1}$  as a function of momentum $k$ for the $\eta=\bar 1$ bands. Close to zero momentum, $k=0$, the $S^x$ component has opposite signs in the topological and trivial phases. 
The parameters are the same as in Fig. 2 of the main text. 
}
  \label{fig:Fig6}
\end{figure}

In Fig. \ref{fig:Fig6}, we present results for the $S^x(k)$ and $S^y(k)$ components of the bulk quasiparticle spin as a function of momentum $k$ for the $\eta=\bar 1$ bands. As discussed in the main text, the $S^x$ reverses its sign as the system undergoes the topological phase transition.
It is also interesting to note that ${\bf S}_{\lambda\eta}(k=0)$ is strictly aligned along the external magnetic field, see Fig. \ref{fig:Fig6}. 

\begin{figure}[b]
\centering
\includegraphics[width=8cm]{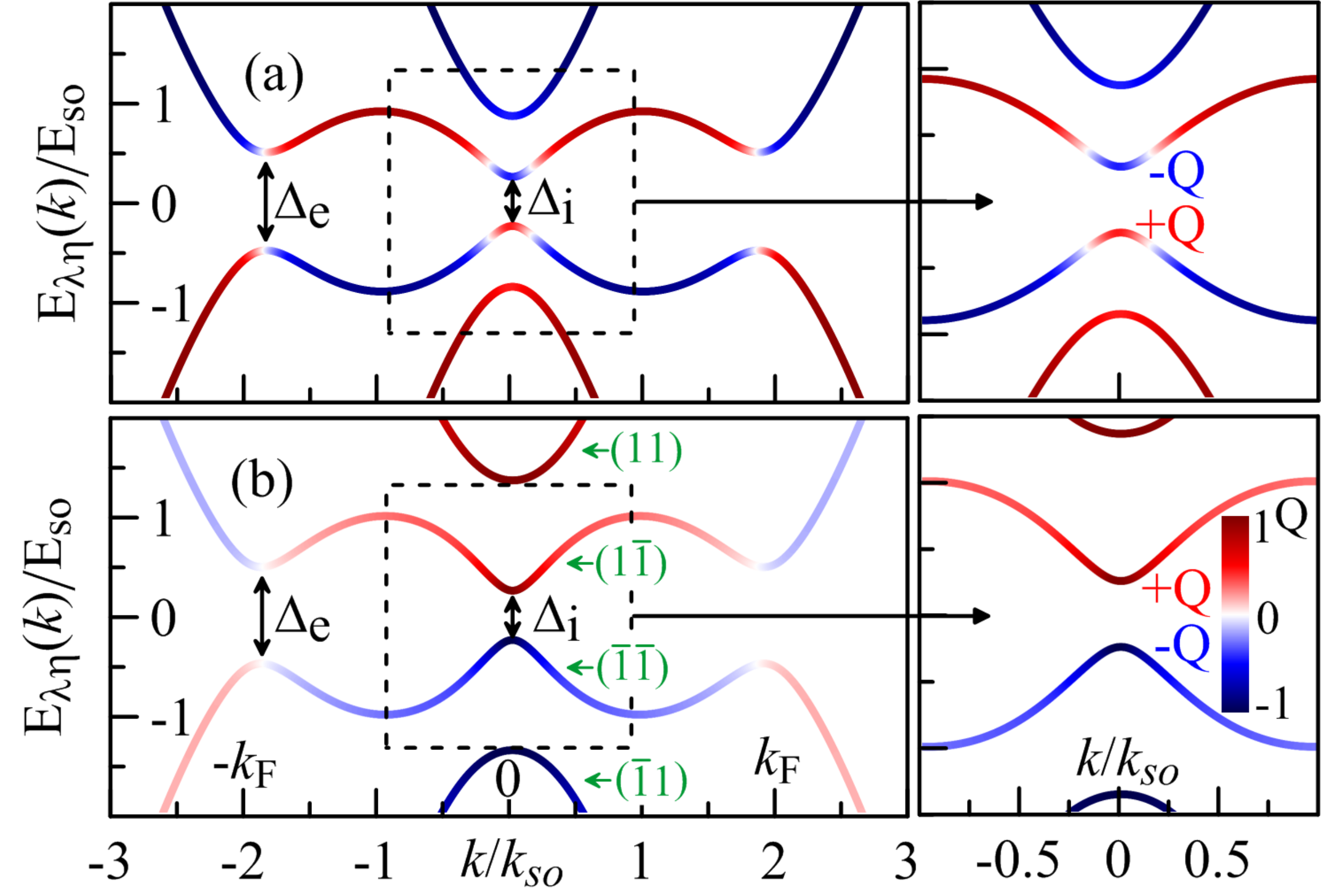}
\caption{The same plot as in Fig. 2 of the main text but for the charge $Q_{\lambda \bar 1}$ instead of the spin projection $S^x_{\lambda \bar 1}$. Again,  the nanowire is in (a) the trivial and (b) topological phases. The blue/ red color corresponds to the negative/positive sign of the charge $Q$. Around $k=0$,  the charge of quasiparticles with fixed momentum $k$ belonging to the lowest energy band $\eta=\bar 1$ in the trivial phase is opposite to the one in the topological one. The insets show the part of the spectrum around $k=0$ for which such the reversal of the sign of $Q$ occurs. We use the following parameters $\mu=- 0.25E_{so}$, $\Delta_{sc}=0.5E_{so}$ and $\Delta_{Z}= 0.3E_{so}$ ($\Delta_{Z}=0.8E_{so}$) in the trivial (topological) phase.
  }
  \label{fig:Fig8}
\end{figure}

\section{Quasiparticle charge }\label{charge_results}

In this Appendix, we present the results on the bulk quasiparticle charge $Q_{\lambda \eta} (k)$, following the line of discussion for  $S^x_{\lambda \eta} (k)$ in the main text. In Fig. \ref{fig:Fig8}, we again plot  the band structure of the system in (a) the trivial and (b) in the topological  phase as was done in Fig. 2 of the main text. However, this time we indicate with blue/red colors negative/positive sign of the quasiparticle charge $Q_{\lambda \eta} (k)$. Again, we focus on the area around $k=0$, where the inversion of the topological bulk gap takes place.  The reversal of the sign of $Q_{\lambda \bar 1} (k)$ indicates that the system went through a topological phase transition. 

\begin{figure}[ht]
\centering
\includegraphics[width=8cm]{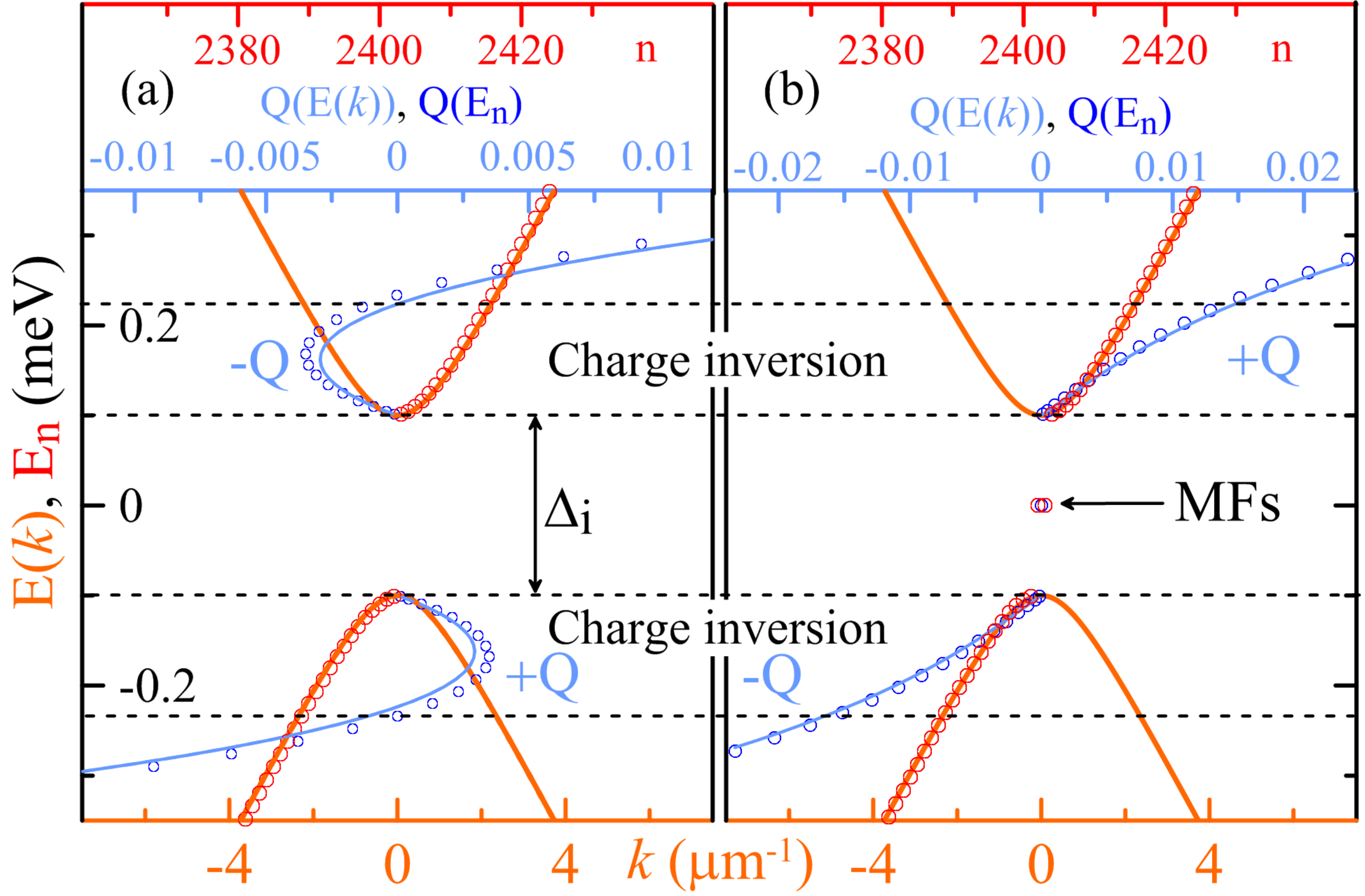}
\caption{ The same plot as in Fig. 5 of the main text for $\mu=0$ but for the charge $Q_{\lambda \bar 1}$ instead of the spin projection $S^x_{\lambda \bar 1}$.
The energy spectrum $E_n$ (red circles) and charge $Q_n$ (blue circles) calculated  for the open system (b) is in good agreement with  $E(k)$ (orange solid line) and $Q(E(k))$ (blue solid line) obtained for the closed system (a).
The sign of the charge around $k=0$ reverses as the system goes through the topological phase transition.}
  \label{fig:Fig10}
\end{figure}

Similarly to Fig. 5 of the main text, we compare results obtained for closed (momentum space) and open (tight-binding model) systems, see Fig. \ref{fig:Fig10}. 
Also for the quasiparticle charge $Q$, we find very good agreement between the two methods. Indeed,  close to $k=0$, $Q$ reverses its sign as the system goes from the topological to the trivial phase.

\section{Disorder}

This section is devoted to address different types of disorder in more detail. For example, in Fig. \ref{fig:Fig12}, we add to our model randomly-oriented magnetic impurities characterized by $\delta \Delta^{i}_{Z,j}$ at the site $j$. Here, we consider a random Zeeman field that can occur in all three spin directions, so $i=x,y,z$. The reversal of the sign of the spin component along the external magnetic field takes place even in
 the presence of such magnetic disorder.
 
% we have plotted the same plot as the Fig. 5 of the main text but for a magnetic disorder this time. The magnetic disorder is induced via an onsite perturbation of the magnetic field in all the directions $\delta\vec{\Delta}^{i}_{Z,j}$ with $j$ corresponding to the site $j$ and $i=x,y,z$ the directions in space.

\begin{figure}[ht]
\centering
\includegraphics[width=8cm]{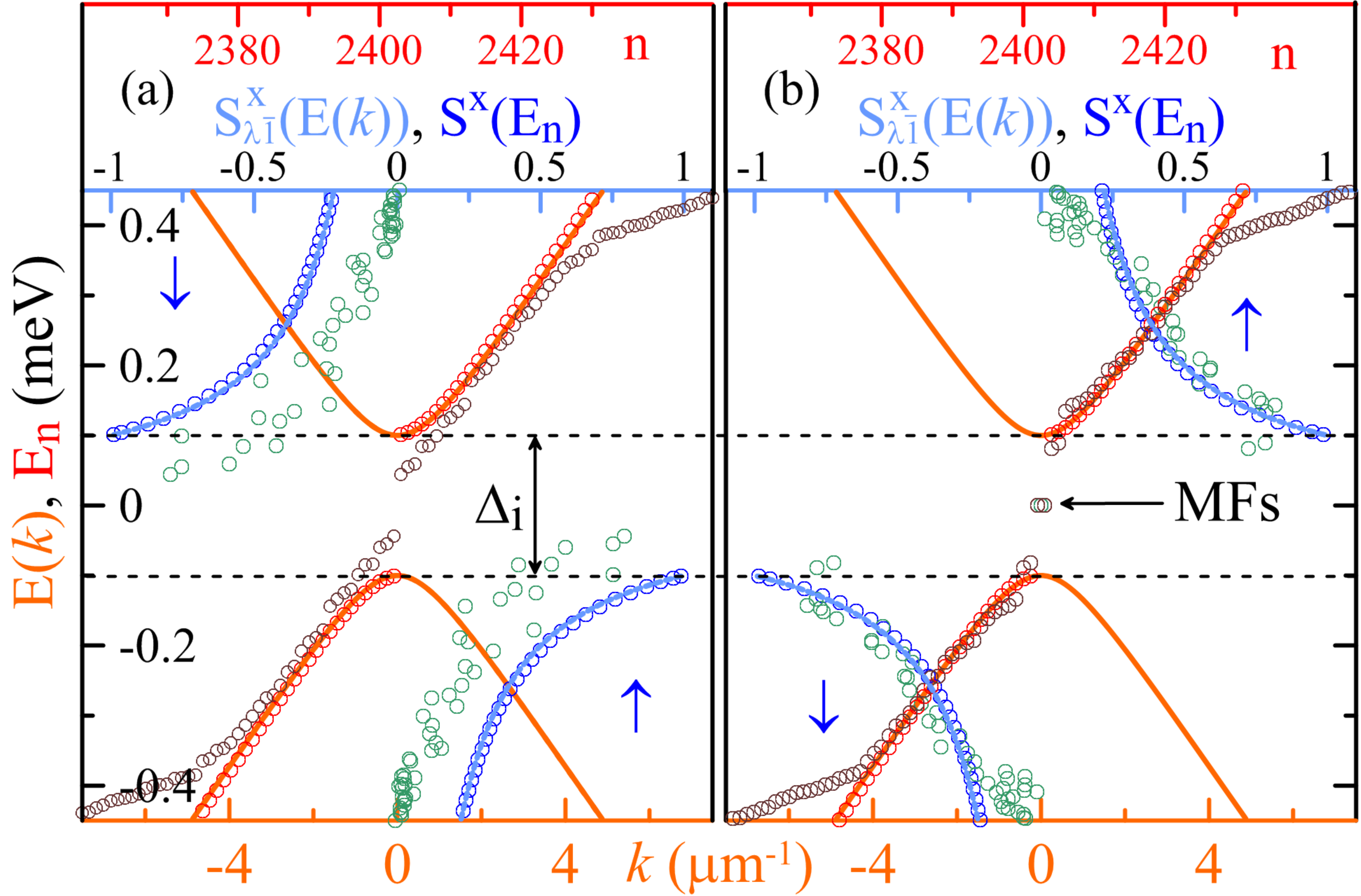}
\caption{ The same plot as in Fig. 5 of the main text but instead of potential disorder we introduce magnetic disorder of the strength $|\delta\Delta^{i}_{Z,j}|<0.5meV$. The reversal of the spin sign is present also in this case and can be used as a tool to distinguish between topological and trivial phases.}
  \label{fig:Fig12}
\end{figure}

\begin{figure}[ht]
\centering
\includegraphics[width=18cm]{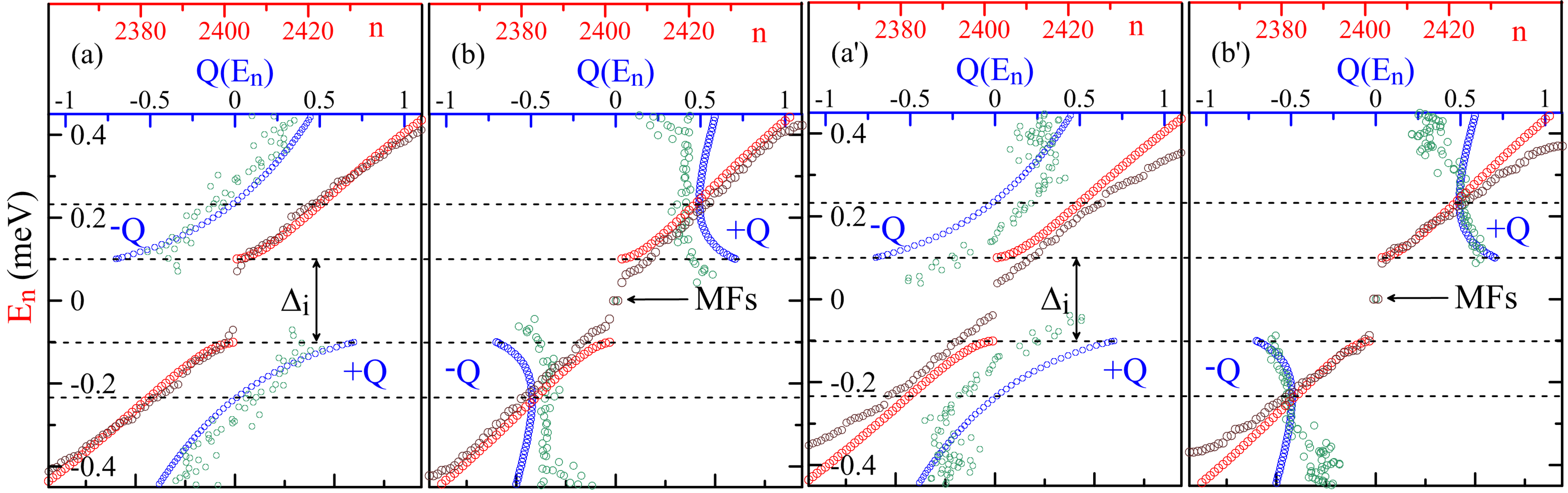}
\caption{ The same plot as in Fig. 5 of the main text but instead of spin we focus now on charge.  The chemical potential is fixed to $\mu=-0.7$ meV and we consider both (a-a') trivial phase with $\Delta_Z= 0.6$ meV and (b-b') topological phase with  $\Delta_Z=0.8$ meV. The energy spectrum $E_n$ (red circles) and charge $Q_n$ (blue circles) are calculated  for the open system.
In addition, we consider in panels (a,b) a random on-site disorder potential of strength $|\delta\mu_j\vert <$ 1meV. In panels (a'b') we consider a random onsite magnetic disorder with $|\delta\Delta^{i}_{Z,j}|<0.5meV$ for $i=x, y, z$.
The charge $Q(E_n)$ (green circles) corresponding to $E_n$ (brown circles) of the disordered nanowire undergoes the same reversal of the charge sign as in the clean case, demonstrating its robustness.}
  \label{fig:Fig13}
\end{figure}

To have a complete overview of disorder effects, in Fig. \ref{fig:Fig13} (a),(b) and Fig. \ref{fig:Fig13} (a'),(b') we study, respectively, the effect of the chemical potential disorder and of the magnetic disorder on the inversion of charge. Again, we confirm that our results are quite robust against disorder.

%\begin{figure}[t]
%\centering
%\includegraphics[width=8cm]{charge_numerics_disorder_B_05meV.pdf}
%\caption{ The same plot as in the previous Fig. \ref{fig:Fig13} but for onsite magnetic disorder $|\delta\Delta^{i}_{Z,j}|<0.5meV$ for $i=x, y, z$.}
%  \label{fig:Fig14}
%\end{figure}

\section{Multisubband system}

\begin{figure}[ht]
\centering
\includegraphics[width=8.6cm]{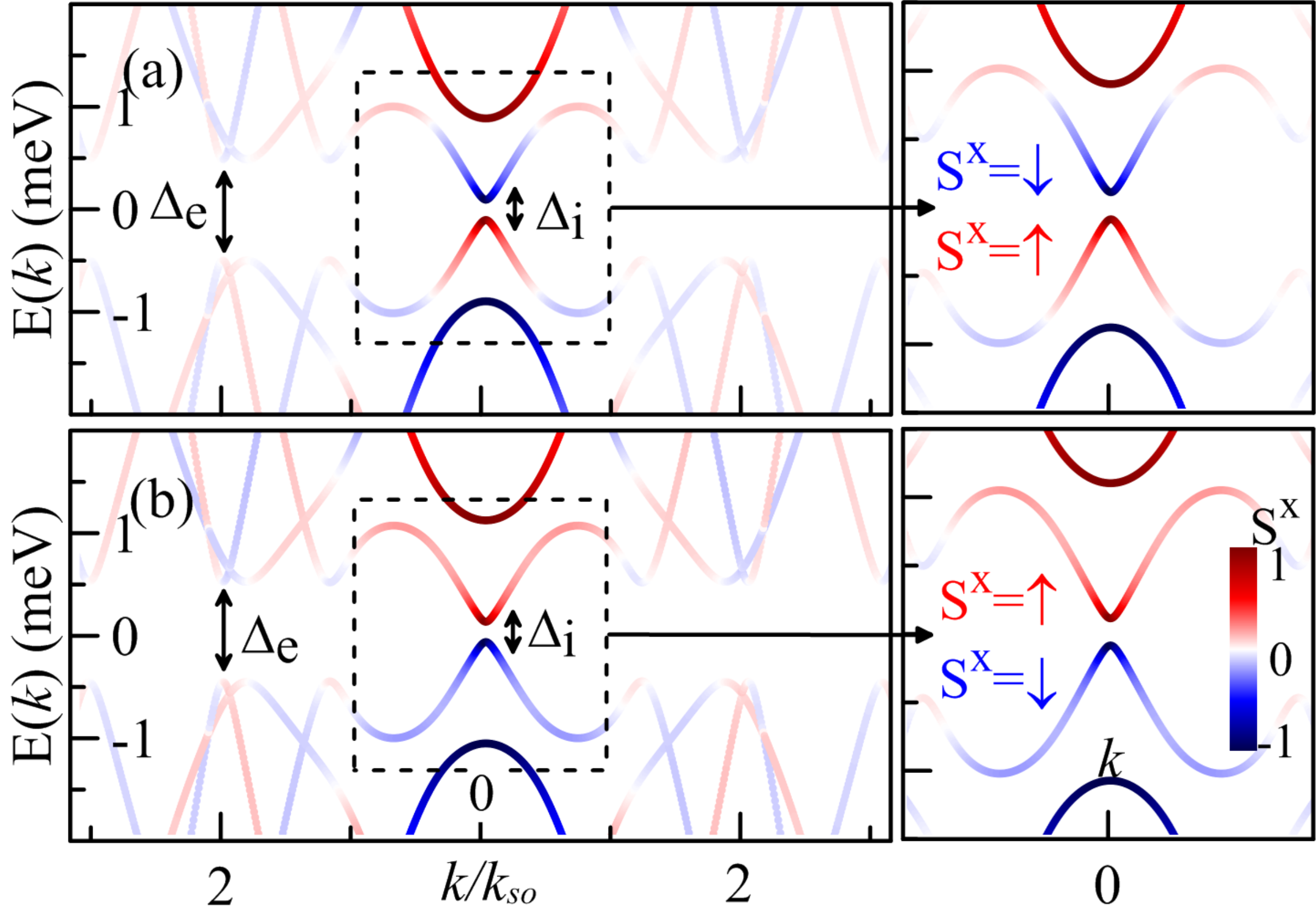}
\caption{The same plot as in Fig. 2 of the main text but for the system with three subbands. The results are obtained for parameters $\Delta_{sc}=0.5$ meV, $t_y$=0.4t, $\alpha_y=0.2\alpha$  (a) in the trivial phase $\Delta_{Z}=0.4$ meV and (b) in the topological phase $\Delta_{Z}=0.6$ meV. One can clearly see that the spin inversion of $S^x$ associated with the band inversion around $k=0$ is not affected by the presence of the other subbands.}
  \label{fig:Fig11}
\end{figure}

The presence of several subbands in the system can, in principle, play a role in the physics studied in this work. To show that our results apply also to multi-subband nanowires, we study numerically a nanowire with three ($N_y=3$) filled subbands, where we also take into account the transverse Rashba SOI of the strength $\alpha_y$. The system is described by the Hamiltonian $H$ in the extended basis of wavefunction $\Psi_{k,j}$, where we have added a label $j$ to distinguish sites in the transverse direction (in our example, there are three sites in the transverse direction),
% $H = \sum_{k} \Psi_k^\dagger\mathcal{H}_k \Psi_k$ 
% defined as
%$H_m~=~\sum_k\Psi_{k,l}^\dag\mathcal H_m(k)\Psi_{k,j}$ where:
\begin{align} 
&H=\sum_{k} \sum_{j=1}^{N_y} \Psi_{k,j}^\dag \Big\{ [-2t\cos(ka)-\tilde{\mu}]\tau_z +\Delta_{sc}\tau_x+\Delta_{Z}\sigma_x+2\tilde{\alpha}\sin(ka)\sigma_y\tau_z\Big\}\Psi_{k,j} \nonumber\\
&\hspace{30pt}+\sum_{k} \sum_{j=1}^{N_y-1}\Big\{\Psi_{k,j+1}^\dag[-t_y-i\alpha_y\sigma_x]\tau_z\Psi_{k,j}+H.c.\Big\}.
\end{align}
Here, $t_y$ is the hopping matrix element in the $y$ direction. The corresponding spectrum as well as the spin polarization of the bands is shown in Fig. \ref{fig:Fig11}. Again, we clearly observe the reversal of the spin at the topological  phase transition point. This allows us to conclude that our results are also valid for multisubband nanowires.

\section{Spin density}
In this subsection we focus on the spatial profile of the $x$-component of the spin density $S^x(j)$, see Fig. 7. The gap inversion occurs at $k=0$, thus the corresponding wavefunctions are slowly oscillating. As a result, in spite of the fact that $S^x(j)$ depends on the position $j$ along the wire, we observe that the sign of $S^x(j)$ stays constant almost over the whole wire length. As a result, the inversion of the spin component can be detected locally with an STM tip at almost any position.
%for the first few states above the gap for the system in the topological and trivial phases for the system parameters same as for the Fig. 5. One can expect to realize tomography of the spin density for these state by the spin polarized STM tip.

\begin{figure}[ht]
\centering
\includegraphics[width=8.6cm]{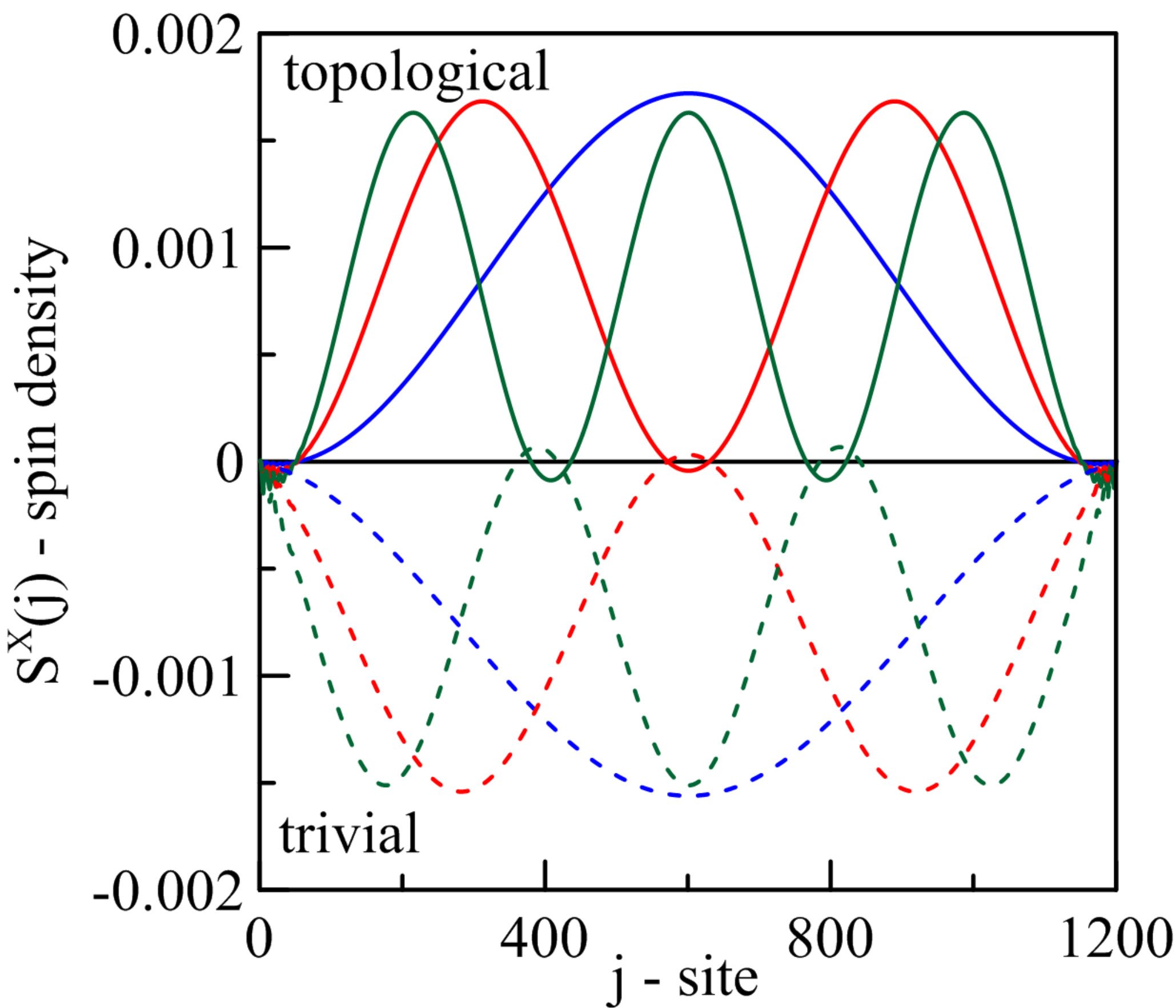}
\caption{The $x$-component of the spin density $S^x(j)$ for the first (blue), second (red), and third (green) states above the gap in topological (solid line) and trivial (dashed line) phases. Importantly, the sign of $S^x(j)$ is almost position-independent. As a result, one clearly observes the spin inversion at any given site $j$ for almost all sites along the wire.
The parameters are the same as in Fig 5. of the main text.}
\label{fig:Fig12}
\end{figure}

\end{document}